\documentclass[
 reprint,
 superscriptaddress,
 amsmath,amssymb,
 aps,
]{revtex4-1}

\usepackage{graphicx}
\usepackage{dcolumn}
\usepackage{bm}
\usepackage{hyperref}

\usepackage{listings}
\usepackage{color}
\usepackage{subfigure}

\definecolor{dkgreen}{rgb}{0,0.6,0}
\definecolor{gray}{rgb}{0.5,0.5,0.5}
\definecolor{mauve}{rgb}{0.58,0,0.82}

\begin{document}

\title{QRunes: High-Level Language for Quantum-Classical Hybrid Programming}

\author{Zhao-Yun Chen}
 \email{czy@originqc.com}
 \affiliation{Key Laboratory of Quantum Information, CAS}
 \affiliation{Origin Quantum Computing, Hefei, China}
 
\author{Guo-Ping Guo}
 \email{gpguo@ustc.edu.cn} 
 \affiliation{Key Laboratory of Quantum Information, CAS}

\date{\today}

\begin{abstract}

Hybrid quantum-classical algorithms have drawn much attention because of their potential to realize the "quantum advantage" in noisy, intermediate-scale quantum (NISQ) devices. Here we introduce QRunes, a cross-platform quantum language for hybrid programming. QRunes can be compiled to various host backends, allowing the user to write portable quantum subprograms. The hybrid programming is based on the type system, which is used to decide where a statement should be run. We also introduce Qurator, a VSCode plugin that has QRunes language support and two host backends.

\end{abstract} 
\maketitle

\section{Introduction}
Quantum computing has drawn much attention in recent years, and many prototypes of quantum computers have been made\cite{kok_linear_2007, schindler_quantum_2013, barends_coherent_2013, li_controlled_2018}. The quantum advantage may be achieved in the near future\cite{NISQ_preskill_quantum_2018}, which means quantum computers could outperform classical computers for certain problems\cite{Grover, shor_polynomial-time_1999, hhl_algorithm_2009}. Many algorithms have been written that seek the quantum advantage, such as boson sampling\cite{clifford_classical_2017} or classical-quantum hybrid algorithms for quantum chemistry\cite{peruzzo_variational_2014, omalley_scalable_2016, kandala_hardware-efficient_2017} and clustering\cite{weinstein_analyzing_2013}.

Quantum language was first believed to be the appropriate programming language for the quantum computer. The original idea was that a code written in quantum language would be compiled into a quantum instruction set or assembly language, directly controlling the quantum computer. However, near-term quantum computers are not likely to run independently. Quantum hardware is considered an accelerator for the classical computer. In real use, classical-based quantum programming interfaces are more appealing because they more easily handle quantum and classical counterparts.

On the basis of these programming interfaces, we have designed QRunes, which is a cross-platform, high-level quantum programming language. A program written in QRunes is compiled into a user-specified target, which can be any programming interface or quantum instruction set. The compiled target is written in the classical language, with each function defined as callable. QRunes provides a "script" that allows the user to write the code, prepare the environment and invoke the compiled function.

QRunes supports enough logical features in quantum programming. Besides some basic features like quantum gates, measurement and packing a series of gates into a function, QRunes supports definition of variables, selections, loops, quantum control flows and operations on the classical memory of the quantum computer.

In its first release, QRunes collaborates with Qurator, a VSCode plugin, supporting two backends: QPanda written in C++ and pyQPanda written in Python. After the program is compiled into the C++ or Python target, the program can be run with the help of the native C++ compiler or Python interpreter. Qurator also helps display the results if they are written in standard output in Json format.

\section{Motivation for Developing QRunes}
Before introducing QRunes, it is necessary to clarify our motivation. There are already many quantum languages or software packages for quantum computing\cite{van_tonder_lambda_2004, green_quipper:_2013, green_introduction_2013, javadiabhari_scaffcc:_2015, noauthor_scaffold:_nodate, wecker_liqui|>:_2014, quil_2016, QSI_2017,ibm_qasm_2017, Q_sharp_2018, khammassi_cqasm_2018}, including QPanda developed by our team. Why should we develop a new quantum language?

As a background, we consider the following two problems. One is that a quantum programming environment should meet the demand of hybrid quantum computing. Another is the reusability of quantum codes.

\subparagraph{Hybrid Quantum Programming}

Hybrid quantum programming combines the power of the quantum and classical computers. This involves two scenarios: Quantum control flows and variational quantum algorithms (VQAs).

Quantum control flows are related to the feedback control of a quantum program. A measurement operation maps the result to a classical variable. A feedback control means the following program is decided by the result of the variable. Written in pseudocode, the procedure has this form:
\begin{lstlisting}[language=C++,numbers=none]
Measure q -> c;
if (c == 0){
    X(q);
}
else{
    Y(q);
}
\end{lstlisting}
If the qubit \emph{q} is measured to be 0, an X gate follows the qubit \emph{q}, otherwise a Y gate. Similarly, a quantum version of \emph{while} can also be defined. 

\begin{figure}[htbp]
    \includegraphics[width=0.45\textwidth]{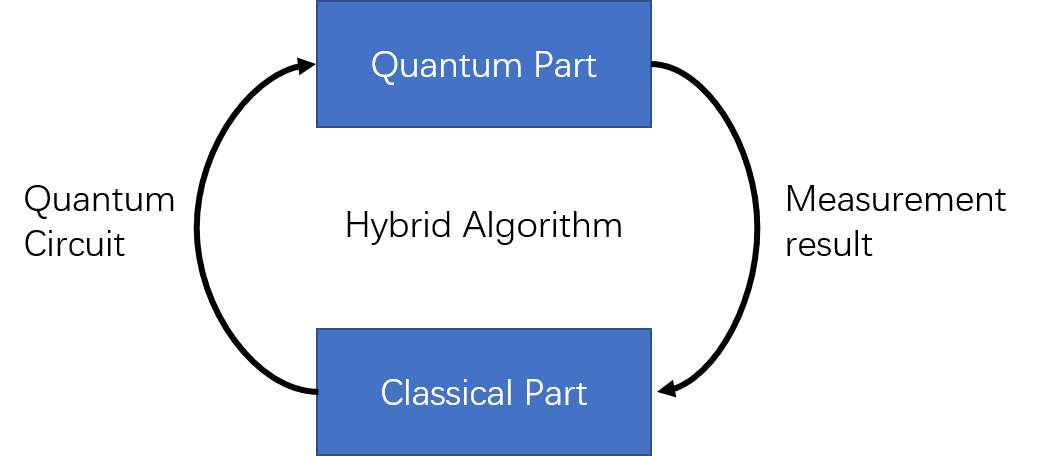}
    \caption{Hybrid quantum programs}    
    \label{Hybrid}
\end{figure}

The VQA has drawn much attention and is likely to be used in quantum computers in the near term\cite{qaoa_2014, qcl_2018, havlicek_supervised_2018, QGAN_PRA_2018, q_autoencodrs, q_BoltzmannMachine, farhi_classification_2018, schuld_quantum_2018, schuld_circuit-centric_2018, liu_differentiable_2018}. A VQA (such as the loop in Fig. \ref{Hybrid}) requires a customized classical-quantum feedback loop to optimize the loss function for the unknown variables, as these variables usually appear in quantum circuits.

These scenarios require us to think about the variables related to the classical computer and how they are compiled in the whole algorithmic process, instead of only thinking about the construction of quantum circuits.

\subparagraph{Reuse of Quantum Codes}
In classical programming, cross-platform languages are widely used and convenient in modern software engineering. Quantum codes, implementing the same algorithms, are written differently and in different languages, programming architectures and interfaces. Therefore they are not cross-platform.  Reusable, or portable, quantum subprograms omit language differences and implementation details, and only contain the logic of the quantum algorithms.

\subparagraph{Design Principles of QRunes}

QRunes does not introduce any new feature in quantum computing, nor is it meant to be compiled into a concrete quantum instruction set. QRunes is designed for writing portable quantum subprograms and is targeted to any existing quantum programming interface. We extract the universal features of quantum-classical hybrid programs and convert the code to the target.

To face the demand of building a hybrid program, we have to be careful with all types of variables. We design a system that has three primary types: \emph{quantum}, \emph{classical} and \emph{assist-classical}. These types correspond to the hardware layer where the variables, expressions and statements are handled. For detailed information, see Section \ref{TypeVariable}.

QRunes does not work independently; instead, it depends highly on the features of the compiling target. It does not include quantum machine configuration, such as initialization, program loading and running and fetching of results. QRunes only focuses on the subprogram that only runs in the quantum computer. The configuration is handled by the external environment, which is the compiling target itself.

\section{Introducing QRunes}

A QRunes program consists of three main parts: \emph{Settings}, \emph{QCode} and \emph{Script}.

\subparagraph{Settings}
This part defines the global settings for the compiler. All settings are written in the form
\begin{lstlisting}[numbers=none]
    SETTING = ATTRIBUTE;
\end{lstlisting}
There are two frequently used settings: \emph{language} and \emph{autoimport}. Language is used to specify the compiling target and can be set to Python or C++. If "autoimport = True", the preset snippets are pasted onto the top of the converted code and are meant to import (include) and configure the environment.

\subparagraph{QCode}
QCode is the fundamental part of QRunes. In this part, the user builds functions to control the quantum computer. QCode is eventually compiled into the selected target and can be invoked by classical language with the generated application program interface (API). Next, we will introduce the language design and syntax of QCode. 

\subparagraph{Script}
Script is optional. This part writes the wrapper code for the generated QCode.

In general, compiling a QRunes file generates a header and a .cpp file (selecting C++ as the target) or a Python package (selecting Python as the target). These files can be then imported into the classical codes or used in \emph{Script}.

\subsection{Basic QCode Syntax}
Basically, QCode consists of the following parts:
\begin{itemize}
\item Include
\item Variable Declaration and Initialization
\item Quantum Selection and iteration
\item Classical Control Flow
\item Function Definition
\item Statement
\item Comment
\end{itemize}
QRunes is a quantum language that incorporates grammar from many classical programming languages. The modulization of a QRunes program is based on the function definition and function calls. As in many other languages, the statement is the basic unit in the function definition.

Here is a simple example of the QCode of a QRunes program:
\begin{lstlisting}[language=C++,numbers=none]
foo(qvec q, cvec c, int s){
    let size = len(q);
    for (i = 0 : size){
        if (i != s){
            CNOT(q[i],q[s]);
        }
    }
    MeasureAll(q,c);
}
\end{lstlisting} 

\subsubsection{Comment}
The comment in the QRunes is identical to that in the C programming language. Code wrapped by /* and */ forms a block comment, and // comments a line.

\subsubsection{Variable Declaration and Definition}
Variable declarations without definition only appear in the parameter list of the function. The types of variables are described in Section \ref{TypeVariable}.

Variable definition is a kind of statement. The \emph{let} statement can be used to define and initialize an assist-classical variable (for more about assist-classical variables, please refer to Section \ref{AssistClassicalVariable}). QRunes has both a quantum data type, which refers to qubits, and classical data type to refer to classical memory.

\subsubsection{Function Definition and Function Call}
A function in QRunes is defined as a representation of quantum subprograms. A function definition consists of the function name, parameter list and function body. The function body has a list of statements. A function does not need a return type, and return is not defined either. Calling a function means sequentially performing every statement of the quantum operation type (see Section \ref{QuantumOperation}) in the function.

Because QRunes highly depends on the target quantum programming architecture, the actual behavior of defining and calling a function is based on the features of that architecture. The script of an external program can call the compiled function with the same name as the function in the QCode. When QPanda is set as the target, any function in QRunes will be compiled into a C++ function returning a QProg. An empty QProg is defined at the beginning of the function, and all quantum operations are sequentially inserted into it.

\subsubsection{Statement}
The statement is the basic unit for the function body. Quantum and assist-classical are two types of statements, which are introduced in Section \ref{StatementType}. The statement syntax is similar to that of the C language.

\subsection{Type System of QRunes}\label{TypeVariable}
\begin{figure}[htbp]
    \includegraphics[width=0.45\textwidth]{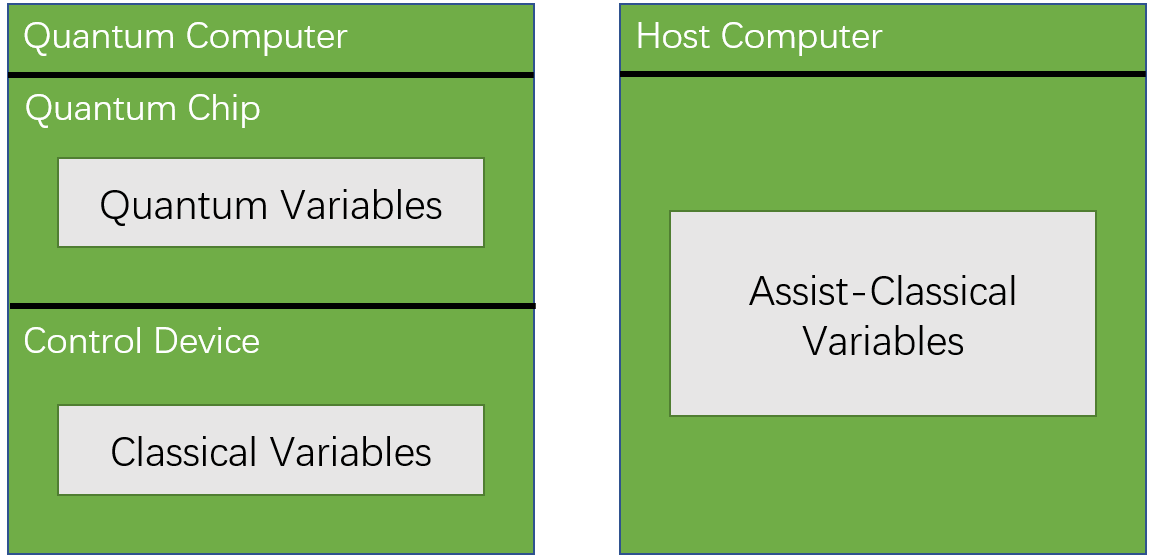}
    \caption{Places where different variables are handled.}
\end{figure}
\subsubsection{Quantum Type}
The quantum type is the way to describe one qubit or an array of qubits. We now support qubit and qvec, which denote a qubit and a qubit vector, respectively. The variable of a quantum type is a mapping to the qubit in the quantum device, and is directly used in the quantum gates:
\begin{lstlisting}[language=C++,numbers=none]
H(q);	    // q is a qubit.
H(qs[0]); // qs is a qvec.
\end{lstlisting}
Another common use is as the passing parameter in the quantum function:
\begin{lstlisting}[language=C++,numbers=none]
foo(qubit q, qvec qs){
   /* use q and qs in quantum gates 
      or function calls. */
}
\end{lstlisting}
The first qubit creation is handled by the host language. All quantum codes only handle quantum operations. We designed them this way because quantum devices in the near term will only have a limited amount of qubits, so there is little meaning in considering dynamic qubit allocation or release.

There is a constraint in using the quantum type, which is that the value of the qubit cannot be changed at any time. We only considered the variable as an address of a qubit with an invisible state that can only be manipulated by the quantum gates, rather than a kind of memory holding qubit data such as amplitude, phase or binary value. The copy of the quantum type is the reference copy, not the value state. An assignment is the only case in which a quantum type appears in an expression. The semantics of an assignment expression is to create a reference or an alias from the right value:
\begin{lstlisting}[language=C++,numbers=none]
qubit alias = q;        // create an alias.
H(alias);				// same as H(q).
qubit q = qs[0];
qvec qs1 = qs[0:3];    
\end{lstlisting}
The expression len(qvec) returns an assist-classical type that represents the vector size. This can be realized because the lengths of all qvec types are determined in the compilation of the quantum program.

\subsubsection{Assist-Classical Type}\label{AssistClassicalVariable}
Assist-classical type describes the classical variables that assist in constructing the quantum program. These variables are determined before the quantum program is transferred to the quantum device. The assist-classical type is not formally considered in the QBuilder system. Instead, the processing of these variables is maintained by the host language or programming interfaces. Here we only deal with the common numerical type and array type. For example,
\begin{lstlisting}[language=C++,numbers=none]
let i = 0;
H(q[i]);    
\end{lstlisting}
The assignment of variable i is handled by the host language and is not transferred to the quantum computer.
Another example is iterating over all qubits and performing Hadamard gates:
\begin{lstlisting}[language=C++,numbers=none]
for (i = 0 : len(q))
    H(q[i]);     
\end{lstlisting}
If the host language supports a dynamic type system (such as JavaScript), or supports a type inference or implicit type (such as C++ and Python), then we only need to use “let” to define an assist-classical type. An example is as follows:
\begin{lstlisting}[language=C++,numbers=none]
let i = 0;
let b = 0.1;
RX(q[i],b);        
\end{lstlisting}
If the type must be explicit, the assist-classical type follows the definition rule of the host language, with a qualifier “host”, for example:
\begin{lstlisting}[language=C++,numbers=none]
host int i = 0;
host double b = 0.1;
// host TYPE identifier = Expression; 
RX(q[i],b);
\end{lstlisting}

\subsubsection{Classical Type}
Classical type handles the classical bits or variables in the quantum computer. In the physical realization, the quantum chip does not work alone, as the control and measurement signals are processed by a set of embedded devices. The measurement results are temporarily saved in the internal memory of these devices. This design is meant to realize the feedback control of the quantum program, such as qif and qwhile\cite{ying_foundations_2016}. A qubit is fragile and will collapse in nano- to microseconds, which is much shorter than the time needed for the qubit system to perform a feedback. Using the embedded devices to realize the control system of the quantum chip meets the feedback demand. 

A classical type is a specialized way to represent data in an embedded system. An example is CBit, which is used to save the measurement result of a qubit. CVec is a CBit array:
\begin{lstlisting}[language=C++,numbers=none]
Measure(q,c);        // perform a measurement.
for (i = 0:len(q)) // measuring a qvec to a cvec.
    measure(q[i],c[i])
\end{lstlisting}
Extracting the classical type is usually related to the driver interfaces of the embedded system that transfer the data from the embedded system to the host computer. This can be handled by the host language. Similarly to the quantum type, the classical type is considered a mapping to the embedded system memory, and the value cannot be modified directly by assignment. Besides, the allocation and release are not considered part of QBuilder.
A classical type can appear in the feedback control sequence of the quantum program, such as quantum selection and iterations. This part will be introduced in the section on quantum-classical feedback.
A classical type can appear in arithmetic expressions. An expression related to a variable of the classical type that defines a calculation in the quantum computer, which is processed in the embedded system mentioned above. This type of expression, which we call classical expression, has an equivalent position to that of the quantum gate. Handling an expression with classical type is different from the way an assist-classical expression is handled, which will be introduced in the section on expression and statement type. The return value of the classical expression is a classical-type variable. A binary operation over both classical and assist-classical types is valid, with the return value also being a classical type.
Here is an example of using a variable:
\begin{lstlisting}[language=C++,numbers=none]
C1 = C2;
C2 = !C2;
Qif (C2) { // do something }
Qif (C1) { // C1 is the negative of C2 }
\end{lstlisting}
Assigning a constant to a classical type is also valid. We have to be careful because the assignment does not actually run before the program is transferred into the quantum computer. In turn, an assist-classical type cannot be assigned by a classical type:
\begin{lstlisting}[language=C++,numbers=none]
C1 = True
Qif (C1) { // do something }

let a = C1; // Bad assignment
\end{lstlisting}

\subsection{Expression and Statement}
\label{StatementType}
We must clarify the expression and statement mainly because performing a quantum program involves three components: the classical computer, control devices and quantum chip. The quantum, classical and assist-classical programs are running in these three components, respectively. The compiling and running times may also be different.
\subsubsection{Quantum Operation Type}
\label{QuantumOperation}
A quantum operation includes the elementary quantum operation and quantum control flow.
\paragraph{Elementary Quantum Operation}
The elementary quantum operation  includes measurement, function calls and classical expression. The quantum gate, measurement and classical expression correspond with the instructions that run in the quantum computer. Translating the QRunes program into the QPanda system has this form:
\begin{lstlisting}[language=C++,numbers=none]
/* a program fragment,
   the comment shows the converted program. */

// prog<<H(q); 
// (prog is defined as a QProg automatically)
H(q);  
Measure(q,c) // prog<<Measure(q,c);
c = c + 1;    // prog<<(c=c+1); ()
\end{lstlisting}
In the last line, c=c+1 returns ClassicalCondition, which represents a symbolic variable with an add and an assignment operation. The ClassicalCondition variable can be further implicitly converted to ClassicalProg and then insert into a quantum program.

The function definition is different from that in the classical language. For a detailed introduction see the subsection on function. The function call is also an elementary operation in the quantum program.

\paragraph{Quantum Control Flow}
The quantum control flow usually denotes the qif and qwhile subprograms. A qif program selects the branch by evaluating the classical expression. A program sample is:
\begin{lstlisting}[language=C++,numbers=none]
Measure(q, c);
qif(c){
    H(q);
} qelse {
    NOT(q);
}
\end{lstlisting}
In this program, if \emph{c} is measured to be 1, a Hadamard is applied to \emph{q}, which will result in a $|0\rangle-|1\rangle$ state on the qubit\emph{q}. If \emph{c} is measured to be 0, a NOT gate is applied, resulting in $|1\rangle$ on the qubit \emph{q}.
The qwhile program is similar:
\begin{lstlisting}[language=C++,numbers=none]
// suppose q is set to |0>+|1>
// c and temp are classical variable
Measure(q, c);
temp = 0;
qwhile(c){
    H(q);
    Measure(c);
    temp += 1;
}
\end{lstlisting}
This program repeats Hadamard and measurement operations on \emph{q}, and uses a classical variable \emph{temp} to record how many times it loops.

To construct a quantum control flow program, the condition and the subprograms (\emph{qif}, \emph{qelse} and \emph{qwhile}) should be assigned. Selecting the branching depends on the classical expression, which is only evaluated in the embedded system mentioned above.
\begin{table*}[htbp]
    \caption{Difference between classical and quantum control flow}
\begin{tabular}{m{5cm}|m{6cm}|m{6cm}}
    & Classical Control Flow & Quantum Control Flow \\
    \hline
  Condition Type & Assist-Classical Expression & Classical Expression \\
    \hline
  Evaluate the condition & In Classical computer before transferring the quantum program & In embedded system after running \\
    \hline
  Existence of the subprogram & Exist only if the condition is true & Exist without aware of the condition is true or not \\
    \hline
  Scope inheritance & Inherite all assist-classical variables & Do not inherite any assist-classical variables \\
    \hline
\end{tabular}
\label{Difference}
\end{table*}

\paragraph{Scoping rule for the quantum control flow}
Scoping is common in modern programming languages and is used to mark the valid place to reference an identifier. It is related to the lifetime of a variable. Most of the time, the scope inherits the symbols in its parent scope. In QRunes, the inheritance of the scope is valid in most parts of the program, except the quantum control flow. Consider a program like:
\begin{lstlisting}[language=C++,numbers=none]
let ac = 1;
measure(q, c)
qif (c==1) {
  // should ac be run in the quantum device?
  ac += 1;     
}
\end{lstlisting}
The expression about \emph{ac}is evaluated depending on the measurement result of \emph{q}, which implies the expression is a classical expression. However, according to the definition above, \emph{ac} is an assist-classical variable, hence the expression is an assist-classical one, which is a contradiction.

There is a special scoping rule based on the rule for quantum control flow statements. The assist-classical variables are invisible to the statement in the body of the quantum control flow, but the quantum and classical variables are visible. Here is an example:
\begin{lstlisting}[language=C++,numbers=none]
let ac = 1;
measure(q, c)
qif (c==1) {
// ac += 1;    
//  bad expression, ac will not be inherited.
let ac = 100;  //  ok
let i = 0;
while (i<ac){
H(qs[i]); // qs is inherited in qif scope.
/* i is inherited normally 
   in the classical control flow. */
i+=1;   
    }
} 
\end{lstlisting}

This prevents an assist-classical expression from being determined after the quantum program is compiled. Table \ref{Difference} shows the difference between the classical and quantum control flows.

\subsubsection{Assist-Classical Operation Type}

Assist-classical is related only to the assist-classical variables and constants. Any expression or statement is evaluated while the program is compiled. Assist-classical design is independent from that of the classical. Consider the following pseudo-code of a quantum program:
\begin{lstlisting}[language=C++,numbers=none]
for (i = 0 to n-1)
    H(q[i]);
\end{lstlisting}
There are two compiling demands. One is to unroll the loop:
\begin{lstlisting}[language=C++,numbers=none]
// for (i = 0 to n-1)
//    H(q[i]);

// Compiling into a simple quantum instruction set.
H 0
H 1
H 2
...
\end{lstlisting}
This is easy and natural for a small-scale quantum computer, because performing a loop in the controlling system is rather difficult. The unrolled program is then converted into the microwave sequence to control the quantum computer, which drops the existence of the variable i and the loop. The function calls are also handled in this way: unrolling all the jumps and branching before running.

This design would be invalid if a large-scale or even intermediate-scale quantum computer were realized. If there are hundreds of qubits, (and performing Hadamard gates on all qubits is common in most algorithms,) it needs too much memory to save the control sequence. Although there has been a design to realize sequential decoding and wave conversion, unrolling all the loops sacrifices too much space, which finally causes unscalability.

Suppose a slightly complex instruction set for the quantum computer that compiles the above program:
\begin{lstlisting}[language=C++,numbers=none]
// for (i = 0 to n-1)
//    H(q[i]);

// Compiling into a simple quantum instruction set.
mov reg1 0 
ld reg2 [n] // load the range into the variable.
ld qreg [q]      // load the qubit from the address q.
LABEL:
add qreg, qreg, reg1 // add the offset.
H qreg          // perform a Hadamard gate.
/* compare the value in reg1
   and reg2 and jumps if true. */
jcmp LABEL, reg1, reg2 
\end{lstlisting}
Compared with forcibly unrolling the program, this instruction set allows a constant length of a for-loop program. This set is much more difficult to implement, and the latency for execution of quantum gates may not be negligible.

On account of the two demands of compiling, we separate the definitions of assist-classical and classical, which correspond to an unrolling plan and complex instruction set, respectively. The program in QRunes is written as follows:
\begin{lstlisting}[language=C++,numbers=none]
for (i = 0 : n)
// i is assist-classical, unroll the for-loop.
  H(q[i]);  

// c is a classical variable.
c = 0;
qwhile (c < n) {
// c is evaluated in the embedded device.
    H(q[c]);    
    c += 1;
}
\end{lstlisting}
 
\paragraph{Assist-Classical Expression}
QRunes supports fundamental binary and unary operators, which refers to the grammar of the C language:
\begin{lstlisting}[language=C++,numbers=none]
let c = 1;
c += 1; // evaluate to 2.
c == 1; // evaulate to false.
\end{lstlisting}

Similar to the assist-classical definition, users can use the expression in the native host language:
\begin{lstlisting}[language=C++,numbers=none]
host int c = 1;
host{
    c += 1;
    /* if the host language support 
       ** to denote the power. */
    c **= 2; 
}
\end{lstlisting}

\paragraph{Classical Control Flow}
Classical control flow represents the selection and iteration statement executed in the host computer. The classical control flow involves "if", "while" and "for", with similar meanings in modern languages.

\subsection{Function}
The function is also very common in modern languages. In most cases, a function consists of the returning type, function name, parameter list and function body. The function body is a mixture of statements.

Functions in QRunes are a little different. First, a function does not have a returning type. Instead, the function implicitly returns a quantum program and can be treated as a user-defined quantum gate in the function call:
\begin{lstlisting}[language=C++,numbers=none]
foo(int n, qubit q)
{
    H(q[n]);
}

...

bar(qubit q){
    let n = 1;
    foo(q, n); // identical to H(q[n]).
}
\end{lstlisting}
Not only can a function be called by another quantum function, but it can also be called by the host language in the script. This is how the classical language invokes Qcodes within QRunes. Here is an example of a full quantum program written in QRunes:

\begin{lstlisting}[language=Python,numbers=none]
// the qcode part
@qcodes:
Test(qubit q, cbit c, cbit temp)
{
    temp = 0;
    H(q);
    measure(q,c);
    qwhile (c) {
        temp += 1;
        H(q);
        measure(q,c);
    }
}

@script:
from pyqpanda import *
init()
q0 = qAlloc()
c = cAlloc()
temp = cAlloc()

result = direcly_run(Test(q0,c,temp))
print(temp.eval())
print(result)

finalize()
\end{lstlisting}

This program executes a quantum subprogram, which performs a while loop. In the loop a Hadamard gate is repeatedly applied to the qubit and the measurement. For every measurement the classical variable \emph{c} has a 50\% probability of being true, which adds 1 to the counter-classical variable\emph{temp}. After the program is run, the results of \emph{temp} and \emph{c} are printed. Variable \emph{c} is always "False", representing the end of the loop. Variable \emph{temp} records how many times the loop is performed, which has an average value of 1.

\section{Qurator: VSCode Plugin for QRunes}

VSCode is a light, strong, open-source code editor developed by Microsoft. Extension Marketplace provides many plugins that have supported the code-editing, debugging and linting of almost all mainstream programming languages.

The QRunes language support provided by Qurator makes quantum programming more user-friendly.

\subparagraph{File Template}

Qurator provides a file template for QRunes that has the form:
\begin{lstlisting}[language=C++,numbers=none]

// declare a constant m and initialized as 0.908.
let m = 0.908;

// Function Declaration.
simple_test(qvec q,cvec c);

// Function Definition.
simple_test(qvec q,cvec c){
    H(q[0]);
    RX(q[0],m);
    Measure(q[0],c[0])
}
\end{lstlisting} 
This is meant to help a new user build a quantum program.
    
\subparagraph{Auto-Completion}
Auto-completion is the most basic function of the language support. Qurator provides an intelligent auto-completion code by listing all the valid symbols in the corresponding scope.

\subparagraph{Verification}
Verification is another fundamental function. Qurator automatically checks the grammar and displays error messages in the output box.

\subparagraph{Hovering}
When the mouse hovers over a word in the program, Qurator provides information about it.

\subparagraph{Highlight}
Keywords are colored by the Qurator, which helps the readability of the program.

\subparagraph{External Language Support}
Qurator supports QCode collaboration with an external program written in the supported classical language. In writing the script, Qurator uses third-party VSCode plugins to provide the language support. This is controlled by "language" in the settings.

\subparagraph{Launching and Displaying a Result}
Qurator enables the user to write a full executable program in one QRunes file. For example, if Python is chosen and the user writes some script to run and fetch a result, Qurator performs the full launching process, including compiling and running, and finally displays the result in another tab. In the result panel, a user can click the result to display the data histogram.

\section{Conclusion}
We have introduced QRunes, a high-level programming language suitable for composing portable quantum subprograms and hybrid quantum-classical algorithms. The grammar and syntax rules are mainly drawn from the C language. We have defined a type system to clarify the compiling rules of QRunes. We have quantum-type, classical-type and assist-classical-type variables, expressions and statements. These types are processed with different rules, and they are expected to be run in the quantum computer, control system and host computer, respectively. 

QRunes is not directly compiled into any quantum hardware or instruction set; instead, it is designed to be "cross-platform" and can be compiled to a target quantum programming architecture. QRunes thus applies its settings to instruct the compiler with information on the target, and uses its script to enable the quantum codes to collaborate with the classical host. 

In the first release, QRunes has two quantum architecture backends: QPanda and pyQPanda, and works with the VSCode plugin Qurator. Qurator provides QRunes language support, including highlighting, auto-completion, syntax checking, launching and result displaying. In future releases, QRunes will support more quantum programming architectures and extend more grammar rules. 

\bibliography{references}

\begin{appendix}
    
\section{Introduction to QPanda}

Quantum Programming Architecture for NISQ device application (QPANDA) is a package of quantum programming interfaces developed by Origin Quantum. QPanda is mainly written in C++, and has both C++ and Python APIs. The Python version (pyQPanda) can be downloaded from PyPI (using “pip install pyqpanda”). QPanda is now open-source in GitHub.

In QPanda, we designed a framework to construct multi-level quantum operations, such as a quantum program, quantum circuit, quantum gates and control flows. The user can make a quantum circuit/program by inserting the quantum operations into it.
\subparagraph{Example 1. A program to build a Bell state and perform many-shot measurements}
$\newline$
\begin{lstlisting}[language=C++,numbers=none]
#include "QPanda.h"
using namespace QPanda;
map<string, size_t> Bell_State()
{
    QuantumMachine *machine = initQuantumMachine(
          QuantumMachine_type::CPU_SINGLE_THREAD
        );
    auto qubit_list = machine->Allocate_Qubits(2);
    auto cbit_list = machine->Allocate_CBits(2);
    auto qprog = QProg();
    qprog << H(qubit_list[0]) 
          << CNOT(qubit_list[0], qubit_list[1])
          << MeasureAll(qubit_list,cbit_list);
    auto resultMap = runWithConfiguration(qprog, cbit_list, 100);
    return resultMap;
}    
\end{lstlisting}
QPanda is built classically and is the controller of the quantum computer. Besides the APIs for constructing a pure quantum program, QPanda also includes the interfaces to input data to / gather data from the quantum computer. These interfaces are all combined in the header file "QPanda.h". After including the header, controlling a quantum device is as simple as using a C++ library:
\begin{lstlisting}[language=C++,numbers=none]
QuantumMachine *machine = initQuantumMachine(
    QuantumMachine_type::CPU_SINGLE_THREAD
);
\end{lstlisting}      
\begin{figure}[htbp]
    \includegraphics[width=0.45\textwidth]{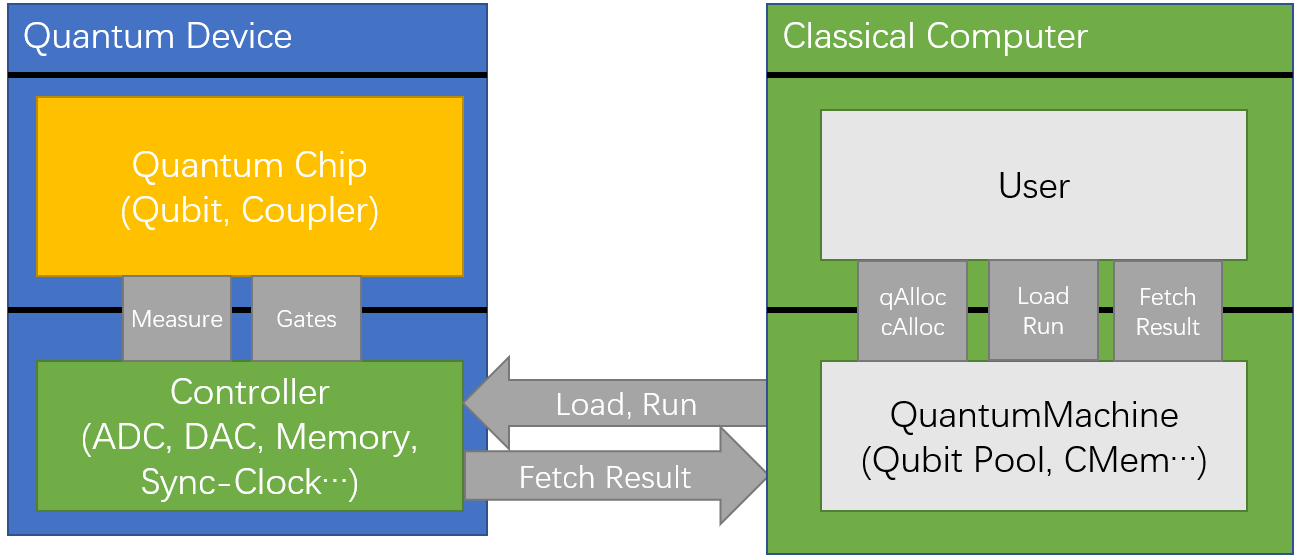}
    \caption{Quantum machine model showing the relation between the quantum device and classical computer.}
\end{figure}
Before we construct a quantum program, a quantum machine must be initialized. QPanda makes an abstraction of all quantum backend types, which constitute the abstract base class QuantumMachine. Within QuantumMachine, a quantum backend is initialized, does qubit/cbit allocation, loads a quantum program and fetches the results. By feeding the argument, the corresponding quantum machine is then returned as a pointer, which can be used to call polymorphism interfaces.

QPanda now supports three types of simulator backends: CPU (with OpenMP multithreading), CPU\_SINGLE\_THREAD (a single-threaded CPU version) and GPU (CUDA implementation). In most cases, the speed of these three backends is in the order CPU\_SINGLE\_THREAD $<$ CPU $<$ GPU.
\begin{lstlisting}[language=C++,numbers=none]
auto qubit_list = machine->Allocate_Qubits(2);
auto cbit_list = machine->Allocate_CBits(2);    
\end{lstlisting} 
Qubit allocations are important before building a quantum program. The system automatically allocates a qubit in the qubit pool and returns a pointer to a virtual qubit in the virtual qubit pool. This design is meant to hide the details of the qubit layout in the quantum computer. All qubits are rearranged before running to validate and optimize the quantum program.
\begin{figure}[htbp]
    \includegraphics[width=0.45\textwidth]{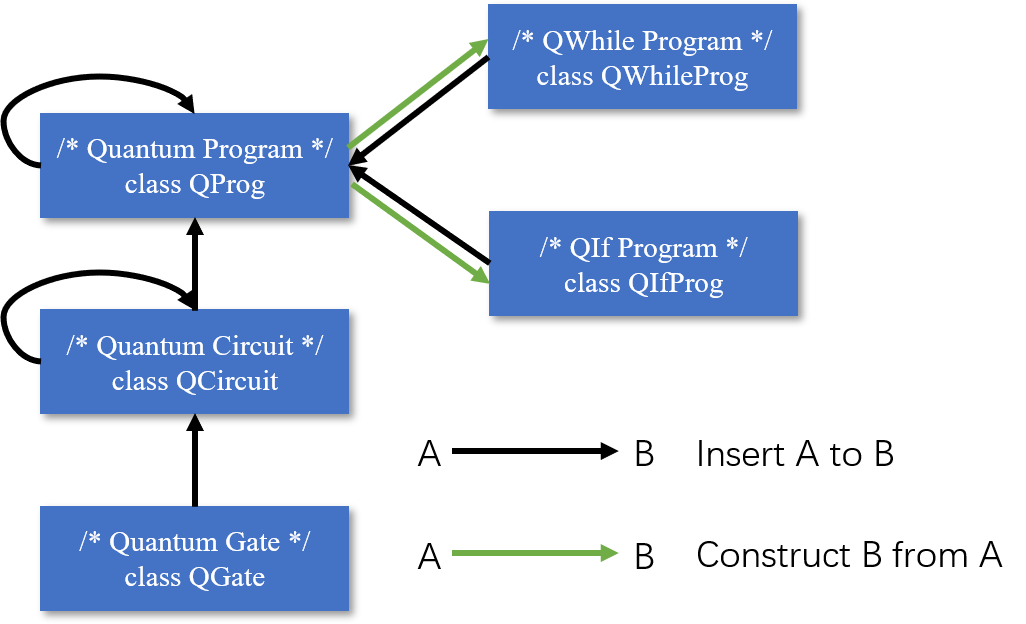}
    \caption{Quantum program components in QPanda. A QGate can be inserted into a QCircuit, and a QCircuit can be inserted into a QProg. Two QCircuit or QProg components can be merged together. QIf/QWhileProg nodes are constructed by a QProg (such as true and false branches in post selection), and they can be inserted into a QProg. The structure enables recursive construction of a quantum program and quantum control flows.}
\end{figure}
If the program runs on a real quantum computer, it is important to allocate cbits. The system also returns a pointer to the CMem, which is connected with the classical memory in the qubit controlling system.
Allocate\_Qubit/Allocate\_CBit allocates one, while an API with a plural form allocates as many qubits/cbits as the parameter gives, which returns an std::vector type:
\begin{lstlisting}[language=C++,numbers=none]
auto qprog = QProg();
qprog << H(qubit_list[0])
      << CNOT(qubit_list[0], qubit_list[1])
      << MeasureAll(qubit_list,cbit_list);     
\end{lstlisting} 

After that, we create an empty quantum program container qprog and insert quantum gates into it. The Bell state can be prepared by sequentially applying a Hadamard gate and CNOT gate. The declarations of the Hadamard and CNOT gates are:
\begin{lstlisting}[language=C++,numbers=none]
QGate H(qubit*);
QGate CNOT(qubit* control, qubit* target);      
\end{lstlisting} 
QPanda overloads the left-shift operator to denote insertion, which is inspired by the stream insertion representation in C++ (std::cout::operator<<).

MeasureAll is a utility function that measures each qubit to the corresponding cbit. The realization is quite simple:
\begin{lstlisting}[language=C++,numbers=none]
using namespace QPanda;
QProg MeasureAll(std::vector<Qubit*> qubits, 
    std::vector<CBit*> cbits)
{
    if (qubits.size() == cbits.size())
    {
      QProg prog = QProg();
      for (int i=0; i<qubits.size(); ++i) 
        prog << Measure(qubits[i], cbits[i]);
      return prog;
    }
    else throw std::invalid_exception(
        "Qubits and cbits must have same sizes");
}       
\end{lstlisting} 

A basic way to encapsulate a quantum circuit is to write a function that returns QProg or QCircuit. The function parameter could be anything, such as an integer to mark the index of the qubit, or a double-precision number to mark rotation angles (just as we have QGate RX(Qubit*, double)). This is how QPanda realizes scalable procedural quantum programming interfaces. 

\begin{lstlisting}[language=C++,numbers=none]
auto resultMap = runWithConfiguration(
    qprog, cbit_list, 100); 
\end{lstlisting} 
Finally, we use the quantum virtual machine to run the quantum program “qprog” many times. The parameters of the function “runWithConfiguration” are the target quantum program, cbit registers and running times.
The result of 100 independent experiments of this quantum program is: resultMap[“00”]=47, resultMap[“11”]=53.
\subparagraph{Example 2. Quantum Control Flow}

\end{appendix}

\end{document}